\documentclass[a4paper,twoside]{article}

\usepackage{epsfig}
\usepackage{subcaption}
\usepackage{calc}
\usepackage{amssymb}
\usepackage{amstext}
\usepackage{amsmath}
\usepackage{amsthm}
\usepackage{multicol}
\usepackage{pslatex}
\usepackage{xcolor}
\usepackage{apalike}
\usepackage{algorithm2e}
\usepackage[bottom]{footmisc}
\usepackage{SCITEPRESS}   

\makeatletter
\let\@fnsymbol\@arabic
\makeatother

\begin{document}

\title{Computing: Looking Back and Moving Forward}

\author{\authorname{Muhammed Golec\sup{1} and Sukhpal Singh Gill\sup{1}}
\affiliation{\sup{1}School of Electronic Engineering and Computer Science, Queen Mary University of London, London, UK}
\email{\{m.golec, s.s.gill\}@qmul.ac.uk}
\thanks{{\textit{\textbf{\textcolor{blue}{Preprint Version Accepted for Publication in Proceedings of the 21st International Conference on Smart Business Technologies (ICSBT 2024), Dijon, France, July 9–11, 2024. DOI: https://dx.doi.org/10.5220/0012855200003764}}}}}{Keynote Paper}
}



\abstract{The Internet and computer commercialization have transformed the computing systems area over the past sixty years, affecting society. Computer systems have evolved to meet diverse social needs thanks to technological advances. The Internet of Things (IoT), cloud computing, fog computing, edge computing, and other emerging paradigms provide new economic and creative potential. Therefore, this article explores and evaluates the elements impacting the advancement of computing platforms, including both long-standing systems and frameworks and more recent innovations like cloud computing, quantum technology, and edge AI. In this article, we examine computing paradigms, domains, and next-generation computing systems to better understand the past, present, and future of computing technologies. This paper provides readers with a comprehensive overview of developments in computing technologies and highlights promising research gaps for the advancement of future computing systems.}

\onecolumn \maketitle \normalsize \setcounter{footnote}{0} \vfill

\section{\uppercase{Introduction}}

Revolutionary developments such as the discovery of fire, the invention of writing, and the printing press have shaped human history. One of these developments is computing technologies, which find a place in almost every field today \cite{gill2024modern}. Finance, where daily transactions are carried out, healthcare, where patient records are kept and analyses are performed, and social media areas can be given as some examples \cite{golec2021ifaasbus}. One of the revolutionary developments in information technology is Artificial Intelligence (AI), which can think like a human and perform certain tasks for now. Because it can be applied to many fields, it has the advantage of being faster and more effective than humans in solving complex problems \cite{iftikhar2023ai}. It is thought that if AI is integrated with quantum computing, another technologically revolutionary development, it can increase the quality of life for humanity in many areas \cite{gill2024quantum}.

\begin{figure*}[t]
	\centering
	\includegraphics[scale=0.60]{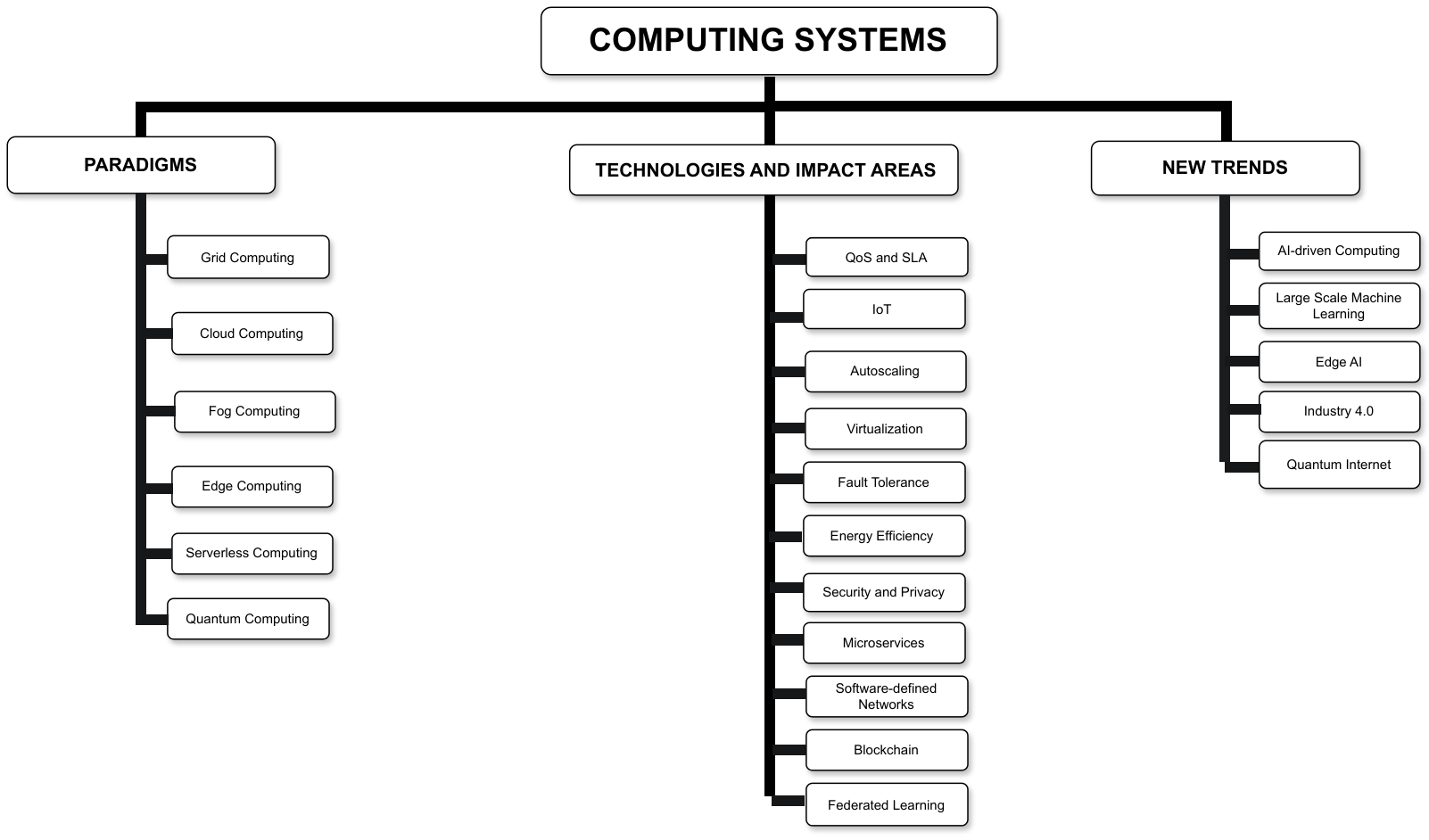}
	\caption{The Overview of Advancements in Computing Systems and Technologies}
	\label{fig:tax}
\end{figure*}

This article gives readers a general perspective by examining the developments in the computing paradigm from the past to the present, current technology applications, trends, and challenges, and next-generation computing concepts. Figure \ref{fig:tax} shows an overview of the advancements in computing systems and technologies. First, existing technologies and domains such as  Internet of Things (IoT), Blockchain, and quantum computing are examined. We then discuss how these technologies shape computing. In the last part, we will examine new generation computing trends and challenges, such as Industry 4.0 and Quantum Internet, which are the future of computing.

\section{Computing Paradigms} 
In this section, we examine the main paradigms to better explain the evolution in computing technology.

\subsection{Grid Computing} 
To complete complicated activities or find solutions to big issues, grid computing is a distributed computing paradigm that permits the sharing and coordinated use of the same networked computing resources \cite{gill2024modern}. Many different fields, such as weather forecasting and the financial industry, utilize it for quick fixes like risk assessments and the examination of massive volumes of data. Several benefits of this paradigm have drawn the interest of both academic and private-sector researchers, including the following \cite{jacob2005introduction}:

\begin{itemize}
    \item \textit{The Ability to Scale:} Resources can be adjusted in size to meet shifting demands and workloads.
    
    \item \textit{The Use of Resources:} Distributed resources are pooled to enhance resource utilization, which lowers the free capacity to maintain optimal efficiency.
    \item \textit{The Economy of Cost:} Using available resources, we can solve tasks requiring a lot of computing at a low cost, eliminating the need for specialized infrastructure.
\end{itemize}

In addition to these advantages, it still brings challenges that need to be solved, such as risks to security, complexity, and variability in performance.

\subsection{Cloud Computing} \label{subsec:Cloud}

In contrast to traditional computing systems, Information Technology (IT) based this paradigm allows computing resources to be pooled and made available over the Internet, and transactions to take place on distant servers \cite{golec2023cold}. Many domains of civil and commercial applications, including web and enterprise applications, have widely adopted cloud computing. Users can benefit in a number of ways, including (i) cost savings through a reduction in hardware and infrastructure expenses, (ii) resource scalability to adapt to variations in demand, and (iii) access from any location with Internet access \cite{murugesan2024neural}. However, it also presents drawbacks such as resource management, privacy, and security issues that require attention.

Three main categories typically classify different cloud delivery models, which offer consumers varying degrees of service \cite{golec2023atom}:

\begin{itemize}
    
\item  \textbf{IaaS (Infrastructure as a Service):} In this service model, consumers rent cloud service provider (CSP) resources such as virtual machines and storage. CSPs are responsible for the server's infrastructure operations, while users are only responsible for infrastructure management.

\item \textbf{PaaS (Platform as a Service):} It's a service paradigm that keeps customers away from infrastructure management so they may concentrate solely on developing applications. A PaaS approach gives users access to the required development environments and software tools.

\item \textbf{ FaaS (Function as a Service):} This completely isolates users from infrastructure and service administration. This relieves users of all infrastructure management responsibilities and allows them to concentrate entirely on developing code. Under this service paradigm, users upload single-function bits of code to the platform using an event-driven architecture.
\end{itemize}

\subsection{Fog Computing} 

With the increase in IoT-based applications, the amount of data that needs to be processed has reached gigantic proportions. Research shows that there will be 24 billion enterprise IoT connections by 2030 \cite{Vailshery}. When cloud systems utilize remote servers, an increase in data processing leads to a decrease in Quality of Service (QoS) parameters like latency \cite{golec2023qos}. Fog computing is emerging as an alternative to cloud computing to solve this problem. Fog computing expedites response times by situating data processing and storage near the data generation sources \cite{iftikhar2022tesco}. Time-sensitive applications and Industrial Internet of Things (IIoT), which process large amounts of data, benefit greatly from this approach. While it offers users benefits like reduced latency and enhanced security by bringing processing power closer to the data source, it also presents challenges like resource constraints and compatibility issues due to the use of heterogeneous and distributed devices.

\subsection{Edge Computing} 

The concept of edge computing emerged to solve problems such as bandwidth and latency in solutions such as cloud computing, where processing power and storage move away from the data source \cite{nandhakumar2024edgeaisim}. For this reason, edge devices such as sensors and smartphones are positioned close to data sources. Today, it has wide usage areas such as IIoT and time-sensitive healthcare applications \cite{10335918}. Edge computing provides advantages such as (i) Reducing latency by moving the processing power closer to the data source, (ii) Reducing unnecessary bandwidth usage because it only transmits data that requires high processing power to the cloud, and (iii) Providing data security because it processes the data locally \cite{iftikhar2022fogdlearner}. In addition to these advantages, it brings challenges such as resource management, security and privacy, and coordinated work arising from the distributed architecture of IoT, which are still waiting to be solved.

The reasons and goals of Fog computing and Edge computing are very similar. For this reason, the two concepts are often confused. When both IT models are examined carefully, it is seen that they differ from each other as follows \cite{buyya2019fog}:

\begin{itemize} 

\item Although both computing models aim to reduce latency and unnecessary bandwidth usage, the main difference between them is related to where data is processed and stored. In fog computing, data processing locations are closer to the data source than in cloud, but further away than in edge computing. Examples are gateways and city data centers. In edge computing, the place of data processing is usually IoT and mobile devices where data is produced.

\item  Fog computing is often used in large-scale systems such as IIoT, where data processing and storage are handled at a middleware. Edge computing has a more distributed approach.

\end{itemize}

\subsection{Serverless Computing} 
Cloud computing has introduced delivery models with various advantages since the day it was first introduced. More information about these models is given in Section \ref{subsec:Cloud}. One of these delivery models is Serverless computing, which isolates server maintenance and infrastructure management from customers and combines Backend as a Service (BaaS) and Function as a Service (FaaS) services \cite{golec2023cold}. In this way, customers can focus only on code development by being isolated from all infrastructure management processes. Serverless computing first emerged in 2014 when Amazon Web Services introduced the Lambda model. This service model uses an event-driven architecture and code developers write their codes as functions. Today, serverless-based applications are used in a wide variety of fields such as finance and education.

It has the following advantages over traditional cloud service models (such as IaaS, and PaaS) \cite{golec2021ifaasbus}:

\begin{itemize}

\item The pay-as-you-go pricing model charges only according to the time resources are used,

\item  The auto-scalability feature allows for the dynamic increase or decrease of computing resources in response to an increasing number of requests,

\item  Ease of management, complete isolation of infrastructure operations from customers.

\end{itemize}

In addition to all these advantages, studies continue in the academy and private sector to solve challenges such as cold start latency, security and privacy, and monitoring and debugging.

\subsection{Quantum Computing}

It is a computing paradigm in quantum physics that works based on the principle of using particles in superposition, called qubits and has recently attracted attention in academia \cite{gill2024quantum}. It has the potential to solve problems that require a long time to be solved in a very short time compared to classical computers where all operations are processed with binary bit (0 and 1) logic \cite{gill2022quantum}. For this reason, it can be used in complex areas such as Cryptography, AI, and Molecular Biology that require high processing power.

Quantum computers have advantages over classical computers such as parallel processing ability, quantum superposition, and quantum teleportation. Thanks to these advantages, it provides advantages such as high speed, parallel processing, and new algorithm discoveries  \cite{gill2024quantumitl}. On the other hand, quantum computers are still in the research phase and must overcome the challenges described below to become widely available \cite{o2007optical}:
 
\begin{itemize}

\item  \textit{High Cost:} Creating a more powerful quantum computer is directly proportional to the number of qubits to be added. However, adding qubits is very costly.

\item \textit{Quantum-Induced Challenges:} Quantum computers, which operate on the principles of quantum mechanics, involve sensitive processes such as ensuring qubit stability and quantum errors. Due to these technical difficulties, quantum computers have not yet been produced on a large scale.

\item \textit{Insulation and Cooling:} Quantum stability and Quantum errors are very quickly affected by environmental factors such as heat and noise. Therefore, insulation and cooling technologies must be developed to provide these conditions.

\end{itemize}

\section{Computing Technologies and Impact Areas}

Computing technologies have made a huge impact in various research fields and transformed the way of working. Some of these areas are examined under subheadings in this section.

\subsection{QoS and SLA} 

\textit{Quality of Service (QoS)} refers to metrics such as latency, throughput and response rate used to show the performance of an application or network service. \textit{Service Level Agreement (SLA)} is an agreement made between a service/application and customers that guarantees the quality and reliability of the service. Computing technologies have various impacts on QoS \& SLA \cite{golec2023qos}.

\begin{itemize}
   
\item  \textit{Security and Privacy:} Computing technologies support SLA with security methods to ensure the security of applications and user privacy.

\item \textit{Efficiency:} Computing technologies can improve QoS parameters with high-capacity processors and infrastructure to increase the performance of applications.

\item  \textit{Application Flexibility:} Computing technologies can make services/applications more flexible by increasing resources to meet demand fluctuations.

\end{itemize}

\subsection{IoT}

All devices that can connect to the Internet and communicate data with each other are called IoT \cite{sen2021advantages}. It has many uses in civilian and military areas, from smart home appliances to autonomous vehicles. Computer technologies have a very wide impact on IoT.

\begin{itemize}

    \item   \textit{Data Analytics:} IoT devices analyze the data collected through sensors and obtain valuable information as a result of this analysis. For example, in an IIoT scenario where production is monitored, predictive maintenance can detect the engine that will fail and thus prevent production disruption.

    \item \textit{Smart Health Applications:} Patient monitoring and disease diagnosis studies are carried out using IoT devices and sensors. In this way, it is aimed to prevent unnecessary health expenses and fatal diseases through early and accurate diagnosis.

    \item \textit{Smart City Applications:} Includes sustainable solutions such as city infrastructure and energy management, such as traffic lights in cities and energy optimization in homes.
    
\end{itemize}

\subsection{Autoscaling}

The concept of autoscaling refers to the automatic increase or decrease of system resources to meet demand fluctuations, which is frequently used in computing technologies \cite{golec2023qos}. Generally, it occurs when launching new instances during demand fluctuations in order to meet SLAs in cloud services. This is also necessary for resource and cost optimization of applications and services.

\subsection{Virtualization} 

It is a computing technology that allows the resources in a system to be used efficiently and divided into different operating systems \cite{chiueh2005survey}. It is especially popular in IaaS-based delivery models provided to users by major cloud providers.

\subsection{Fault Tolerance} 

It refers to measures that can tolerate service interruptions caused by errors that may occur in a system. It includes backup processes called Redundant Array of Independent Disks (RAID) in cloud systems to prevent data loss due to interruptions that may occur in the system \cite{plank1997tutorial,chouikhi2015survey}.

\subsection{Energy Efficiency}

It refers to a system performing the same work with less energy. Values such as energy spent while performing a task and $CO_2$ emissions are critical in reducing environmental pollution \cite{golec2023atom,feng2012survey}. This has led to the proposal of new technological advances such as green IoT and green cloud computing.

\subsection{Security and Privacy} 

Security and privacy are two of the most fundamental problems in IT systems. Security includes comprehensive issues such as unauthorized access to a system and theft or modification of data in the system. Privacy refers to all situations that may jeopardize user privacy, such as biometric data \cite{golec2022aiblock,de2019security}.

\subsection{Microservices}

It refers to a form of architecture designed to make applications more flexible and easier to develop. Essentially, it aims to divide and develop the application into smaller and independent applications \cite{golec2023cold,pallewatta2023placement}. By dividing the application into smaller services, advantages such as debugging and the ability to use different technologies for each scenario are provided.         

\subsection{Software-defined Networks} 

Software-defined Networks (SDN) is a centralized and software-oriented management approach to ensure more efficient management of resources in a network \cite{fernandes2018sdn}. Computing technologies have various impacts on SDNs.

\begin{itemize}
    
\item \textit{Ease of Network Operation:} SDN provides traffic optimization and dynamic resource allocation to increase the operational capability of the network \cite{sarabia2024progressive}.

\item \textit{Network Scaling:} SDN ensures the scalability of the infrastructure to increase the efficiency of network resources \cite{dhadhania2024unleashing}.

\item \textit{QoS and SLA Improvement:} SDN provides optimization in traffic management to improve QoS and SLA in network applications \cite{wang2016qos}.

\end{itemize}

\subsection{Blockchain} 

It is a computing technology that essentially consists of interconnected blocks containing the hash of the previous block and can be thought of as a digital ledger \cite{golec2023blockfaas}. The data in each block is connected to each other with the hash value of the previous block and a new hash value is obtained for the next block. In this way, any change in any data in the blocks affects all blocks, and data immutability is ensured. Blockchain technologies are used in many areas today \cite{doyle2022blockchainbus,reyna2018blockchain}:

\begin{itemize}
    
\item \textit{Financial Sector:} Blockchain provides a decentralized platform for digital assets such as cryptocurrencies (Bitcoin etc.). This eliminates intermediaries in buying and selling transactions, ensuring safe transfers.

\item \textit{Smart Contracts:} These are contracts that are decentralized and kept on the blockchain network \cite{cr2024blockchain}. As a result, all transactions are public. In addition, trust and data security are ensured between the parties by taking advantage of the data immutability advantage provided by Blockchain.

\item \textit{Product Supply:} Quality control is increased by monitoring supply chains with blockchain-based systems and fraud can be reduced with the principle of data immutability \cite{wang2024does}. In this way, confidence in supply chains and commercial activities can be increased.

\item \textit{Identity Authentication and Voting:} With blockchain-based voting systems, user privacy is ensured by storing personal data securely \cite{hossain2024transforming}. Additionally, manipulations in elections held in countries can be prevented.

\end{itemize}

\subsection{Federated Learning} 

Federated Learning (FL) is a Machine Learning (ML) technique used in distributed architectures \cite{li2020review}. While in ML techniques a centralized system is used for model training, in FL distributed devices are used for model training. FL, a field of computing technologies, has a variety of advantages \cite{mammen2021federated}:

\begin{itemize}
 
\item \textit{Security and Privacy:} Since a central system is not used when training FL models, data security and user privacy risks are lessened.

\item \textit{Fast Training:} Model training is faster because FL is trained in distributed systems close to data sources.

\item \textit{Accessibility:} Since FL are trained in distributed systems, model training is also possible in Internet-limited environments.
 
\end{itemize}

\section{Next Generation Computing: Challenges and New Trends }
This section briefly discusses next-generation computing paradigms, their challenges and new trends.

\subsection{AI-driven Computing} 

AI-driven computing is the integration of AI, ML, and Natural Language Processing (NLP) models into computer systems to process big data and produce solutions for complex models \cite{firouzi2022convergence}. It brings many advantages to the system into which it is integrated \cite{gill2022ai}:

\begin{itemize}

\item  \textit{Prediction Performance:} It is important to analyze data in areas such as finance and health and make predictions with high performance. AI-driven computing increases prediction success rates, resulting in a noticeable increase in risk assessments.

\item  \textit{Cost and Efficiency:} AI-driven computing can reduce costs and increase production efficiency by preventing interruptions in production processes, such as predictive maintenance scenarios.

\item  \textit{Opportunities:} With its innovative solutions, AI-driven computing has the potential to create new business areas and expertise.

\end{itemize}

Apart from all these advantages, AI-driven computing also includes challenges such as ethical issues and new algorithms for optimization in complex operations.

\subsection{Large Scale Machine Learning} 

Traditional ML techniques trained on large datasets have low model performance due to factors such as hyperparameter setting, computational power, and large features. As a result, it is necessary to use techniques such as parallel computing for ML models in large data sets.  Large-scale machine learning methods provide advantages such as improving prediction performance and optimizing transaction cost and speed \cite{wang2020survey}. Besides these advantages, challenges such as feature selection and ensuring data integrity are still waiting to be investigated \cite{liu2024towards}.

\subsection{Edge AI} 

AI-based applications are generally processed in centralized systems with high processing power. Edge AI is an approach that proposes to process data close to the data source, such as IoT and smartphones \cite{bibri2024smarter}. This is because it provides advantages such as lower latency, and security, and reduces unnecessary bandwidth usage compared to centralized systems such as the cloud \cite{singh2023edge}. On the other hand, Edge AI systems offer lower processing power and storage space than cloud-based systems. Also, since it has a distributed architecture, it also brings with it challenges such as the maintenance and management of devices and resource management.

\subsection{Industry 4.0}

It is the use of new computing technologies such as AI and IoT by integrating them into production processes to make production more efficient and economical \cite{teoh2021iot}. Predictive maintenance is a good example of an Industry 4.0 application. By using sensors and ML models, errors are detected in advance in predictive maintenance scenarios, and production disruptions are prevented. Likewise, production optimization is achieved through data analysis methods using data received from sensors. Thus, by enhancing service quality, we can promptly respond to customer demands. In addition to all these advantages, it should not be forgotten that Industry 4.0 brings with it problems that still need to be solved, such as cost, cyber security, and integration.

\subsection{Quantum Internet} 
It is an Internet model that works with quantum principles \cite{wehner2018quantum}. It is expected to replace the classical Internet in the future with its higher security and speed. In the Quantum Internet, quantum cryptography is used to ensure security in communication protocols. Compared to the classical Internet, much faster communication speeds are achieved with quantum teleportation techniques. In addition to all these advantages, quantum computing, and the quantum Internet are still in the research phase. And its integration for long-distance communications is still a challenge \cite{li2024survey}.

\section{\uppercase{Conclusions}}
\label{sec:conclusion}

Computing technologies continue to shape human history with the transformations they offer from the past to the present. This effect can be seen everywhere, from healthcare applications like early diagnosis and medical imaging to military vehicles like unmanned planes. Additionally, technologies such as IoT and AI continue to trend with the opportunities they offer in various industries and processes. Quantum technologies, one of the next-generation developments in computing and still in their infancy, have recently attracted attention for their potential to shape the future. However, embracing these next-generation trends and dealing with their challenges will require separate efforts. In this paper, we take an overview of computing technologies from the past to the future and highlight challenges for researchers.

\section*{\uppercase{Acknowledgements}}

Muhammed Golec would express his thanks to the Ministry of Education of the Turkish Republic, for their support and funding.

\bibliographystyle{apalike}
{\small
\bibliography{example}}

\end{document}